\documentclass[12pt,reprint,aps,groupedaddress,showpacs,nofootinbib,prd]{revtex4-2}
\usepackage{amsmath,amssymb,graphicx,amsfonts}
\usepackage{mathrsfs}
\usepackage{color}

\newcommand{\Xb}{\overset{{\scriptscriptstyle 0}}X{\vphantom{\scriptstyle X}}}

\newcommand{\gb}{\overset{\scriptscriptstyle{0}}g{\vphantom{g}}}

\newcommand{\Rb}{\overset{\scriptscriptstyle{0}}R{\vphantom{R}}}

\newcommand{\pb}{\overset{\scriptscriptstyle{0}}\phi{\vphantom{\phi}}}

\newcommand{\rH}{r_{\mathcal{H}}}

\usepackage{hyperref}
\usepackage{setspace}

\begin{document}

\title{Exact theory for the Rezzolla-Zhidenko metric and self-consistent calculation of quasinormal modes}

\author{Arthur G. Suvorov}
\email{arthur.suvorov@tat.uni-tuebingen.de}
\affiliation{Theoretical Astrophysics, IAAT, University of T{\"u}bingen, T{\"u}bingen 72076, Germany}

\author{Sebastian H. V{\"o}lkel}
\affiliation{SISSA, Via Bonomea 265, 34136 Trieste, Italy and INFN Sezione di Trieste}
\affiliation{IFPU-Institute for Fundamental Physics of the Universe, Via Beirut 2, 34014 Trieste, Italy}

\date{\today}

\begin{abstract}
\noindent{A covariant, scalar-tensor gravity is constructed such that the static, spherically symmetric Rezzolla-Zhidenko metric is an exact solution to the theory. The equations describing gravitational perturbations of this spacetime, which represents a generic black hole possessing an arbitrary number of hairs, can then be derived. This allows for a self-consistent study of the associated quasinormal modes. It is shown that mode spectra are tied to not only the non-Einstein parameters in the metric but also to those that appear at the level of the action, and that different branches of the exact theory can, in some cases, predict significantly different oscillation frequencies and damping times. For choices which make the theory appear more like general relativity in some precise sense, we find that a nontrivial Rezzolla-Zhidenko parameter space is permissible under current constraints on fundamental ringdown modes observed by Advanced LIGO.}
\end{abstract}

\maketitle

\section{Introduction}

The recent release of data by Advanced LIGO and Virgo concerning the first half of their third observing run saw the observed binary black-hole (BH) merger count increase by 39 \cite{ligo2020}. The character of gravitational waves (GWs) from these and previous coalescence events, which are broadly categorised into inspiral, merger, and ringdown phases, can be used to place tight constraints on theoretical departures from the theory of general relativity (GR) \cite{gw150914,ligo19,carson20}, amongst other physics \cite{abb19}. The ringdown phase is especially suited to the experimental validation of the classical no-hair theorems, which state that the post-merger object must be a (astrophysically disturbed) Kerr BH \cite{rob75,heus96,gurs15}. In particular, a breakdown of GR in the strong-field regime, as anticipated from various theoretical considerations (such as non-renormalizability \cite{t1974}), would be signalled by the appearance of non-Kerr features in the quasi-normal mode (QNM) spectrum of the newborn, ringing object \cite{Kokkotas:1999bd,Nollert:1999ji,Berti:2009kk}.

A theoretical classification of possible non-Kerr features is, however, quite challenging. If one is interested in studying tensorial GWs associated with ringdown in some particular, non-GR theory, it suffices to first determine the metric structure of permitted black holes\footnote{Some theories even predict the existence of disjoint families of black holes, which complicates this procedure; see, e.g., Ref. \cite{bar13}.} within that theory, and then derive equations describing their response to gravitational perturbations \cite{mcm1,mcm2}. The eigenvalues of the relevant perturbation operator can be tied to the QNM spectrum, with real parts denoting the oscillation frequencies and imaginary parts denoting the inverse of the damping times due to radiation reaction \cite{Berti:2009kk}. This process is far from trivial to carry out in many cases, and often even the first step provides a computational hurdle since finding exact solutions, especially for realistic and rotating objects, is notoriously difficult. One way to circumvent these issues is to instead consider a parameterised spacetime metric, which arises from some theory-agnostic considerations about what might be expected of astrophysical BHs regardless of the particulars of the gravitational action \cite{vig11,joh11,contfrac}. While parameterised approaches have been largely successful in placing tight constraints on departures from GR from GW data \cite{yun16,Volkel:2020daa,car20a}, they are inherently limited because radiation reaction (especially relevant for the imaginary components of the QNMs) cannot be accounted for self-consistently: without a governing set of field equations, backreaction can only be studied approximately.

A particular solution to this \emph{inverse problem} was recently proposed in Ref. \cite{suv20}, where it was found that a covariant, scalar-tensor theory of gravity can be designed around a particular spacetime (see also Refs. \cite{bois00,noj09,noj19}). More specifically, there are a class of mixed scalar-$f(R)$ theories that possess the special property that for any given metric $g$, there exists a function $f$ such that $g$ becomes an exact solution in that theory if the scalar field satisfies a particular constraint equation. In this way an explicit theory can be reverse-engineered around any given parameterised BH metric, such as those considered in Refs. \cite{vig11,joh11,contfrac}, and gravitational disturbances can be studied self-consistently. It is the purpose of this work to exemplify the inverse mechanism developed in Ref. \cite{suv20} by deriving relevant perturbation equations and studying the QNMs, via Wentzel-Kramers-Brillouin (WKB) methods, of a representative class of parameterised BH metrics.

While realistic BHs are expected to rotate, we consider static metrics in this work to provide a first step towards the building of exact and self-consistent ringdown waveforms for parameterised, non-Kerr objects (see also Ref. \cite{yun09}). We additionally focus on axial perturbations, as these are numerically easier to handle. For demonstration purposes, we consider the Rezzolla-Zhidenko (RZ) \cite{contfrac} class of BH metrics, which represent parameterised departures from a Schwarzschild, and thus vacuum GR due to Birkhoff's theorem, description. {The metric coefficients in the RZ family of metrics are built from continued fraction expansions, so as to be able to efficiently represent a large class of non-Schwarzschild BHs with only a few terms \cite{kon20,suv20x}. Moreover, while the RZ family is very general, free parameters appearing within the metric are chosen to abide by certain recurrence relations so that the spacetime possesses a number of algebraically-desirable properties and can withstand existing observational constraints.} We show precisely how the RZ parameters enter into the self-consistent perturbation equations, associated with some particular branch of an exact theory, and influence the QNM spectrum of a hypothetical, non-Schwarzschild BH.

This paper is organised as follows. Section II reviews the particulars of a mixed scalar-$f(R)$ theory, and shows how it can be used to find solutions to the gravitational inverse problem. Section III presents a derivation for the potential functions associated with a Regge-Wheeler-like perturbation scheme. Some specific QNM calculations are then given in Section IV, with emphasis placed on their relationship to the associated Schwarzschild and RZ values obtained from well-motivated approximation schemes. Some discussion is given in Section V.

\section{Mixed scalar-f(R) gravity}

The vacuum action for the mixed scalar-$f(R)$ theory considered in this work reads
\begin{equation} \label{eq:genscalarten}
\mathcal{A} = \kappa \int d^{4} x \sqrt{-g} f \big( F (\phi) R + \mathcal{V}(\phi) - \chi ( \phi ) \nabla_{\alpha} \phi \nabla^{\alpha} \phi \big),
\end{equation}
where $\kappa = \left( 16 \pi G \right)^{-1}$, $G$ is the (bare) Newton constant (which, together with the speed of light, is set to unity throughout for ease of presentation), $R \equiv R_{\mu \nu} g^{\mu \nu}$ denotes the scalar curvature for the metric tensor $g$, and $F$, $\mathcal{V}$, and $\chi$ are functions of the scalar field $\phi$. The theory described by \eqref{eq:genscalarten} is a particular case of a very general class first introduced by Hwang and Noh \cite{hwang05,hwang02}, {which can be ghost-free for various choices} of $f$ \cite{noj19}; necessary conditions for the health of Friedman-Robertson-Walker universes in the theory described by \eqref{eq:genscalarten}, for example, can be deduced from equations (75) and (76) in Ref. \cite{felice10} (see also Sec. IV). When the function $f$ is linear in its argument $X$, where we define
\begin{equation} \label{eq:xeqn}
X \equiv F(\phi) R + \mathcal{V}(\phi) - \chi(\phi) \nabla_{\alpha} \phi \nabla^{\alpha} \phi,
\end{equation}
the action \eqref{eq:genscalarten} reduces to the standard scalar-tensor one in the Jordan frame \cite{fujii,damour92}. Similarly, the $f(R)$ theory of gravity is recovered for constant scalar field and vanishing potential $\mathcal{V}$ \cite{odin05}. For linear $f$ and constant scalar field, the action \eqref{eq:genscalarten} therefore reduces to the Einstein-Hilbert one. 

The equations of motion (EOM) for the metric and scalar fields read \cite{hwang02,hwang05,suv20}
\begin{equation} \label{eq:field1}
\begin{aligned}
0 =& F(\phi) f'(X) R_{\mu \nu} - \frac {f(X)} {2} g_{\mu \nu} + g_{\mu \nu} \square \left[ F(\phi) f'(X) \right] \\
& - \nabla_{\mu} \nabla_{\nu} \left[ F(\phi) f'(X) \right] - \chi(\phi) f'(X) \nabla_{\mu} \phi \nabla_{\nu} \phi
\end{aligned}
\end{equation}
and
\begin{equation} \label{eq:field2}
\begin{aligned}
0 =& f'(X) \Big[ 2 \chi(\phi) \square \phi + \frac {d \chi(\phi)} {d \phi} \nabla_{\alpha} \phi \nabla^{\alpha} \phi \\
&+ R \frac {d F(\phi)} {d \phi} + \frac{ d \mathcal{V}(\phi)} {d \phi} \Big] + 2 \chi(\phi) \nabla^{\alpha} \phi \nabla_{\alpha} f'(X),
\end{aligned}
\end{equation}
respectively. Equations \eqref{eq:field1} and \eqref{eq:field2} define the vacuum field equations for the theory, though matter can be introduced in the usual way via the stress-energy tensor. 

\subsection{Solving the inverse problem}

As demonstrated in Ref. \cite{suv20}, the configuration space of the theory defined by \eqref{eq:genscalarten} is so large that practically any given metric (spherically symmetric or otherwise) can be admitted as an exact solution for some choice of the functions $f, F, \mathcal{V}$, and $\chi$. In particular, given a metric, the dynamics of the scalar field can be constrained in such a way that that particular $g$ and $\phi$ pair form an exact solution to equations \eqref{eq:field1} and \eqref{eq:field2} provided that $f$ belongs to some particular class of functions.

Specifically, given some metric $g$, if there exists a scalar-field solution $\phi$ to the \emph{kinematic constraint equation}
\begin{equation} \label{eq:constraint}
X_{0} = F(\phi) R + \mathcal{V}(\phi) -  \chi(\phi) \nabla_{\alpha} \phi \nabla^{\alpha} \phi ,
\end{equation}
for some constant $X_{0}$, then, for any $f$ such that $f(X_{0}) = f'(X_{0}) = 0$, that particular $g$ is an exact solution to \eqref{eq:field1} and \eqref{eq:field2}. In general, equation \eqref{eq:constraint} must be solved numerically.

As a concrete example, if $\phi$ is chosen such that $X_{0} = 0$, then $g$ is a solution to the field equations for 
\begin{equation} \label{eq:fxtheory}
f(X) = X^{1+\sigma},
\end{equation}
for any $\sigma > 0$.  Theories of the form \eqref{eq:fxtheory} represent a potentially infinitesimal deviation from standard scalar-tensor gravity at the level of the action, and generalise the $f(R)=R^{1+\sigma}$ theories considered by Buchdahl \cite{buch70} and others. These latter theories are of potential astrophysical relevance since they have been successful in explaining the flatness of galactic rotation curves for $\sigma \sim 10^{-6}$ without the invocation of dark matter  \cite{bohm08,sharm20}. Such values for $\sigma$ are also consistent with cosmological constraints coming from primordial nucleosynthesis \cite{clifbar} (see also Ref. \cite{noj17}). For $\sigma \lesssim 10^{-4}$, stellar structure is largely unchanged relative to GR \cite{cap16}, and thus the theory is capable of accommodating massive neutron stars, such as J0740$+$6620 (with mass $M \sim 2.17 M_{\odot}$ \cite{crom20}), for stiff equations of state. {For weakly-varying scalar fields, these features are likely to persist since the theory \eqref{eq:fxtheory} reduces to the Buchdahl one exactly for $\nabla_{\mu} \phi = 0$ and $\mathcal{V} = 0$.} A curious feature of theories with $\sigma \ll 1$ is that the function $f$ is not analytic, which means that one cannot employ Taylor expansions about small scalar curvatures to investigate the Newtonian limit of the theory. Strong-field tests (coming from QNMs, for instance) are therefore especially germane to theories such as \eqref{eq:fxtheory}. {Note, however, that analytic models can also be constructed; for example, the theory with $f(X) = X + \alpha X^2 + \alpha^2 X^3/4$ for any $\alpha \neq 0$ admits exact solutions with $X_{0} = -2 \alpha^{-1}$.}

In any case, we emphasise here that we are not necessarily advocating for this particular theory as a realistic description for gravitational phenomena, {especially since higher-order theories such as \eqref{eq:genscalarten} are generally susceptible to the Ostrogradsky instability (though see Ref. \cite{wood07}).} The techniques presented in this work are only meant to illustrate that self-consistent studies of QNMs for parameterised spacetimes are possible, and that their properties are sensitive to the particulars of the gravitational action, be it \eqref{eq:genscalarten} or otherwise. In particular, the scalar field solution to equation \eqref{eq:constraint} depends on the functional forms of the potential $(F,\mathcal{V})$ and kinetic $(\chi)$ terms, and therefore behaves differently in different branches of the general theory \eqref{eq:genscalarten}. The QNMs are thus sensitive to the particulars of the dynamics in the scalar sector (see Sec. III. B).

\section{Gravitational perturbations}

In general, the geometric response of a background\footnote{Throughout this work, an overhead zero denotes a background term, i.e., an $\mathcal{O}(\delta^{0})$ quantity.} spacetime metric $\gb$ to gravitational disturbances is studied, at the linear level, by introducing a perturbation term $h$ through $g_{\mu \nu} = \gb_{\mu \nu} + \delta h_{\mu \nu}$, where $\delta \ll 1$ is a formal expansion parameter. We focus on vacuum perturbations in this work, so that $h$ is `sourced' only by the background metric. In general, one may decompose $h$ as a sum of harmonics, viz. $h_{\mu \nu} = \sum_{\ell m} h_{\mu \nu, \ell m}$, where each of the terms $h_{\mu \nu, \ell m}$ are chosen such that they respect the angular symmetries of the spacetime (see below). A further decomposition of $h_{\mu \nu, \ell m}$ into axial and polar sectors is possible, $h_{\mu \nu, \ell m} = h^{\text{axial}}_{\mu \nu, \ell m} + h^{\text{polar}}_{\mu \nu, \ell m}$, where the two pieces are defined by how they transform under parity \cite{regwhel,zerilli}. A separation of this sort is particularly useful for static spacetimes, because the axial and polar sectors decouple \cite{pani13} (see also Ref. \cite{yunes08}). While a derivation of the (generalised) Teukolsky equations relevant for rotating solutions in the theory \eqref{eq:genscalarten} is achievable in principle using the Newman-Penrose formalism or its extensions (see Refs. \cite{suv19a,suv19b} for the $f(R)$ gravity case), solving the resulting equations is computationally challenging \cite{pani13}. We therefore restrict our attention to non-rotating solutions.  

\subsection{Axial modes}

To give a concrete example of how the QNMs of a black hole can be self-consistently studied and compared between different theories, we focus on axial perturbations. Axial perturbations have the benefit that scalar degrees of freedom do not couple to the tensor sector \cite{kob12}, which further simplifies the analysis. The background scalar field still plays a role in the behavior of the perturbations however \cite{kob12,cist15} and modifies the form of the Regge-Wheeler potential \cite{regwhel}. Additionally, we note that even if the perturbed scalar field $\delta \phi$ does not couple to the metric, equation \eqref{eq:field2} predicts the existence of scalar GWs, the amplitudes of which may be non-negligible in some cases \cite{wag70}.

Given some astrophysical measurements of axial QNMs, it has been shown that it is possible to reconstruct the spacetime metric using statistical techniques (e.g., the Bayesian methods detailed in Ref. \cite{Volkel:2020daa}) if one assumes some fixed set of EOM. However, metric and action parameters may be intertwined in the sense that the functional structure of the EOM may itself be dependent on the non-Schwarzschild parameters, which complicates this procedure. The full track of the inverse problem is therefore a non-linear one: one attempts to astrophysically reconstruct the metric by matching QNM measurements to eigenvalues that arise from a metric-dependent operator. As a first step towards a general solution of this difficult problem, we show how one can build a theory around a given spacetime and study its axial perturbations self consistently. This could be then be combined with statistical approaches, such as those detailed in \cite{Volkel:2020daa}, in a future study.

As mentioned earlier, we work with static and spherically\footnote{Theories of the form \eqref{eq:genscalarten} generally allow for BHs with topological horizon structures also \cite{hawk72}, though such objects are likely ruled out by electromagnetic observations of reflection spectra in accreting black holes \cite{namp20}.} symmetric spacetimes in this work for simplicity. The \emph{background} line element, in Boyer-Lindquist coordinates $(t,r,\theta,\varphi)$, is taken to be
\begin{equation} \label{eq:statlineel}
ds^2 = - A(r) dt^2 + B(r) dr^2 + r^2 d \theta^2 + r^2 \sin^2\theta d \varphi^2.
\end{equation}
Making use of the Regge-Wheeler gauge \cite{regwhel}, a generic axial perturbation can be written as
\begin{equation}
\hspace{-0.68cm}h^{\text{axial}}_{\mu\nu,\ell m} = \left(
\begin{array}{cccc}
0 & 0 &  h_{0}(t,r) S^{\ell m}_{\theta}(\theta,\varphi) & h_{0}(t,r)S^{\ell m}_{\varphi}(\theta,\varphi) \\
0 & 0 &  h_{1}(t,r)S^{\ell m}_{\theta}(\theta,\varphi)& h_{1}(t,r)S^{\ell m}_{\varphi}(\theta,\varphi) \\
\ast & \ast & 0 & 0 \\
\ast & \ast & 0 & 0  \\
\end{array}
\right),
\end{equation}
where $S^{\ell m}_{\theta}(\theta,\varphi)= -\csc\theta \partial_{\varphi}Y_{\ell m}(\theta,\varphi)$ and $S^{\ell m}_{\varphi}(\theta,\varphi) = \sin\theta \partial_{\theta}Y_{\ell m}(\theta,\varphi)$ for spherical harmonic $Y_{\ell  m}$, and the asterisk denotes a symmetric entry. The (in general complex) QNM frequency $\omega$ is introduced by taking a Fourier transform of the components $h_{0}$ and $h_{1}$ through 
\begin{equation}
h_{0,1}(t,r) = \frac {1} {2 \pi} \int^{\infty}_{-\infty} d \omega e^{- i \omega t} \tilde{h}_{0,1}(\omega,r).
\end{equation}
The explicit dependence on $\omega$ and the tildes will henceforth be dropped for ease of presentation. 

The methodology used to derive the EOM for $h_{0}$ and $h_{1}$ in the mixed scalar-$f(R)$ theory is remarkably similar to that of GR. In particular, the $\theta \varphi$-component of the  $\mathcal{O}(\delta)$-field equations \eqref{eq:field1} allows for a direct expression of $h_{0}$ in terms of $h_{1}$, viz. [$f'(\Xb) \neq 0$]
\begin{equation} \label{eq:h0exp}
\begin{aligned}
h_{0}(r) =& \frac {i} {2 \omega B} \Bigg[ h_{1} A' + 2 A h_{1}' - \frac {A h_{1} B'} {B}  \\
&+ \frac {2 A h_{1} F'(\pb) \pb'} {F(\pb)} + \frac {2 A h_{1} \Xb' f''(\Xb)} {f'(\Xb)} \Bigg],
\end{aligned}
\end{equation} 
where we note that $\Xb$ is a function of radius only for the spacetime described by \eqref{eq:statlineel}. For the special case of $f'(\Xb)= 0$, a similar expression to that of \eqref{eq:h0exp} can be imposed, though without the final term on the second line. We note, however, that both expressions for $h_{0}(r)$ agree for $\Xb'(r) = 0$ (relevant for the inverse problem discussed in Sec. II. A), i.e., when the constraint equation \eqref{eq:constraint} is satisfied.

Moving forward, we introduce the tortoise coordinate $r^{\star}(r) \equiv \int dr \sqrt{\frac {B(r)} {A(r)}}$, and a new variable $Z$ through
\begin{equation} \label{eq:cases1}
h_{1}(r) = r Z(r)\sqrt{ \frac {B(r)} {A(r) f'(\Xb) F(\pb(r))} }
\end{equation}
for $f'(\Xb) \neq 0$ and $\Xb'(r) \neq 0$, and
\begin{equation} \label{eq:cases2}
h_{1}(r) = r Z(r)\sqrt{ \frac {B(r)} {A(r) F(\pb(r))}}
\end{equation}
otherwise, which allows us to rewrite the $r \varphi$-component of the field equations \eqref{eq:field1} as a Schr{\"o}dinger-like equation. Algebraic manipulations eventually lead to the generalised Regge-Wheeler \cite{regwhel} equation in mixed scalar-$f(R)$ theory, which takes the familiar form
\begin{equation} \label{eq:rweqn}
0 = \frac{d^2}{d {r^{\star}}^2} Z + \left[ \omega^2 - V_{\ell}(r) \right] Z.
\end{equation}
In the general case where $f'(\Xb) \neq 0$, the potential $V_{\ell}$ has the lengthy but simple form 
\begin{widetext}
\begin{equation} \label{eq:vgeneral}
V_{\ell}(r) = V_{0,\ell}(r) - \frac {f A} {F f'} + \frac {3 f'' \Xb'} {B f'} \left[ \frac {4 A + r A'} {4 r}  - \frac {A B'} {4 B} + \frac {3 A F' \pb'} {2 F} + \frac {A \Xb''} {2 \Xb'} + \frac{ A \Xb' f''} {4 f'} \right] + \frac {3 f''' A (\Xb')^2} {2 B f'}.
\end{equation}
The term $V_{0,\ell}(r)$, which reads
\begin{equation} \label{eq:vsing}
V_{0,\ell}(r) =  \frac {\ell \left( \ell + 1 \right) A} {r^2} + \frac {3 \left(A B' - B A' \right)} {2 r B^2} + \frac {3 F' \pb'} {B F} \left[ \frac {4A + rA' } {4 r} - \frac {A B'} {4 B} + \frac {A F' \pb'} {4 F} \right] + \frac {3 A (\pb')^2 F''} {2 B F} + \frac {3 A F' \pb''} {2 B F},
\end{equation}
\end{widetext}
is the piece that can be considered to `survive' for $f(\Xb) = f'(\Xb) = 0$ and $\Xb'(r) = 0$ [cf. equations \eqref{eq:cases1} and \eqref{eq:cases2}]. The latter term \eqref{eq:vsing} in particular is the one most relevant for solutions of the inverse problem. In this case, given some background metric, the scalar field is chosen such that the constraint equation \eqref{eq:constraint} is satisfied, and \eqref{eq:vsing} differs from the GR form explicitly through the presence of $\pb$ and $F$ and implicitly through $A$ and $B$. It is worth pointing out that the only $\ell$-dependent part within $V_{\ell}(r)$ comes from the $\ell (\ell + 1) A/r^2$ term, as expected of theories which predict that (helicity-2) gravitational waves travel at the speed of light; in the geometric-optics limit $\ell \rightarrow \infty$, the perturbation equation \eqref{eq:rweqn} reduces to the null geodesic equation (see Ref. \cite{Volkel:2020daa} for a discussion).

Note that we have considered background spacetimes that are vacuum in this work. This is important to mention since, strictly speaking, any given spacetime can also arise as a solution to the Einstein equations exactly if one allows for arbitrary stress-energy tensors. Perturbations can be studied self-consistently in this approach also. However, in this latter instance, it is unlikely that the matter sector will be associated with a physical Lagrangian (e.g., arising from a fluid) or abide by well-motivated energy restrictions (e.g., dominant energy condition) for a generic background \cite{del93}. Such an approach is therefore not entirely physical in the study of disturbed BHs, since it would be difficult to explain the presence of exotic matter following a vacuum merger event, for example. Nevertheless, a self-consistent approach to the study of perturbations of arbitrary spacetimes within GR is formally possible and the EOM are simply given by $\delta R_{\mu \nu} = 0$ if one assumes that the matter sector is unperturbed by the ringing.

This latter scheme, which we refer to as a `GR-like' approximation throughout, could also be used to study the perturbations of a given spacetime \eqref{eq:statlineel}. In the case of constant scalar field, the `GR-like' Regge-Wheeler equation (see, e.g., equations (14)--(16) in Ref. \cite{Volkel:2020daa}) is recovered exactly from \eqref{eq:vgeneral} for either linear $f$ or for those $f$ with the property that $f'(\Xb) = 0$ [expression \eqref{eq:vsing}]. In many cases of interest therefore, one might expect that this and similar schemes provide a fair representation for the QNMs, even if they are inexact (see the discussion in Ref. \cite{Volkel:2020daa} and below). One of the aims of this present work is to provide a quantification for its accuracy, at least in the simple case of a truncated RZ metric (see Sec. IV). 

\subsection{Dependence of QNMs on the underlying theory}

Expressions \eqref{eq:vgeneral} and \eqref{eq:vsing} imply that QNMs are, in general, sensitive to the particular choices of $f$, $F$, $\mathcal{V}$, and $\chi$. As a demonstration, consider \eqref{eq:fxtheory} for the Brans-Dicke choices with vanishing potential, $F(\phi) = \phi$ and $\chi(\phi) = \chi_{0} / \phi$, where the quantity $\chi_{0}$ is akin to the Brans-Dicke coupling constant. The $\chi_{0} \rightarrow \infty$ limit therefore corresponds to a more GR-like theory; when $f$ is linear, the theory reduces to Brans-Dicke and therefore to GR in this limit. In any case, the constraint equation \eqref{eq:constraint} depends on the value of $\chi_{0}$, and therefore the scalar field does too when considering a solution to the inverse problem. Since the axial potentials $V_{\ell}$ depend on $\pb$ they also implicitly depend on $\chi_{0}$, and the QNMs shift depending on the branch of the general theory \eqref{eq:genscalarten} under consideration. 

This result is familiar from the study of Kerr or Schwarzschild BHs, which are known to ringdown differently in different theories of gravity \cite{barsot08,suv19a,suv19b}, sometimes so dramatically that the same BH may be unstable in one theory and not another \cite{card09,gao19} (see below). For the present case in the study of the inverse problem, this implies that ringdown analyses place constraints on the metric behaviour and on the theory simultaneously; the QNMs for any given quantum numbers depend on $\chi_{0}$ and the free parameters within the metric in a non-trivial way. This adds a layer of complication from a data analysis perspective, since there may be a degeneracy between the physical BH parameters and those parameters that appear within the action functional (see, e.g., Ref. \cite{tatt18}).

The dependence of the QNMs on $\chi_{0}$ implies that stability can be theory-dependent. Specifically, the appearance of a deep enough turning point for $V_{\ell} < 0$ (in a `negative gap') implies the existence of bound states (i.e., $\omega^2 < 0$ modes) that lead to (linear) instability \cite{kon08,yoshida16} (see also Ref. \cite{jar20}). On the other hand, if the potential $V_{\ell}$ is positive-definite, then it is well known that the Schr{\"o}dinger equation \eqref{eq:rweqn} admits no bound states. This condition is not necessary however: using the so-called S-deformation technique \cite{kod03}, Kimura \cite{kimura17} has presented evidence that if there exists a continuous and bounded function $S$ such that 
\begin{equation} \label{eq:sdef}
0 = V_{\ell}(r) + S'(r) - S(r)^2,
\end{equation}
then stability is also assured. These techniques are not necessary for our analysis, since we consider cases with sufficiently small deformation parameters and large enough $\chi_{0}$ (i.e., more GR-like values) so that the negative gap is never too deep to allow for bound states. In general though, a given BH may be stable for $\chi_{0}$ values above some threshold value $\chi_{c}$ that depends on the deformation parameters, and unstable for theories with $\chi_{0} < \chi_{c}$. A thorough investigation of the (in)stability of BHs in different branches of the general theory \eqref{eq:genscalarten} will be conducted elsewhere, since the WKB method used here is not well-suited for identifying bound states.

\section{Quasi-normal modes for the Rezzolla-Zhidenko metric}

The key difference between this work and others which have studied the QNMs of parameterised, bottom-up metrics, is that we build an exact theory around the spacetime using the method detailed in Sec. II. A so that \emph{vacuum} perturbations can be modeled self-consistently. In addition to being of theoretical interest in its own right, this allows, in principle, for a quantification of the accuracy of various widely-used approximations involving test fields \cite{Volkel:2019muj,glamp19,silva20} or mild deviations at the level of the EOM \cite{Volkel:2020daa,mcm1,mcm2}. Where appropriate, we keep track of the relative difference between QNMs computed using the approach presented here and that of the `GR-like' approximation discussed in Sec. III. A, where only the first two terms within \eqref{eq:vsing} are kept. We focus on the case of the RZ metric described in Ref. \cite{contfrac} due its algebraic simplicity, though the methods detailed here can be readily applied to more complicated examples.

For our application we consider a subset of the most general RZ metric with non-vanishing parameters $M, \epsilon, a_1$, and $b_1$, because it provides a good balance of flexibility to approximate non-GR black holes \cite{kon20,suv20x}, but at the same time carries a numerically-manageable number of terms. This choice implies that the BHs we study match Schwarzschild exactly at first post-Newtonian order, because we set $a_0=b_0=0$, which can be motivated from Solar system tests \cite{will18}. We note however that such constraints can be bypassed in some alternative theories of gravity, and these extra parameters could be straightforwardly included as well. For this truncated RZ metric, the coefficients $A$ and $B$ defining the line element \eqref{eq:statlineel} take the form
\begin{equation} \label{eq:afun}
A(r) = 1 - \frac {\rH \left( 1 + \epsilon \right)} {r} + \frac {\rH^3 \left( \epsilon + a_{1} \right) } {r^3} - \frac {\rH^4 a_{1}} {r^4} 
\end{equation}
and
\begin{equation} \label{eq:bfun}
B(r) = \frac {\left(1 + \frac{\rH^2 b_{1}} {r^2}  \right)^{2}} {A(r)},
\end{equation}
respectively. In expression \eqref{eq:afun}, $\rH \equiv 2 M / \left(1 + \epsilon\right)$ defines the location of the event horizon for black hole mass $M$, where $\epsilon$, $a_{1}$, and $b_{1}$ are generic (dimensionless) deformation parameters that are, in principle, to be constrained by observations \cite{contfrac}. The metric reduces to the Schwarzschild metric in the limit $\epsilon = a_{1} = b_{1} = 0$. We emphasise again that the absence of $r^{-2}$ terms in the coefficient $A$ ensures that the spacetime automatically respects many post-Newtonian constraints \cite{will18}. 

For the RZ metric defined by expressions \eqref{eq:afun} and \eqref{eq:bfun}, it is straightforward to solve the constraint equation \eqref{eq:constraint} (even when rotation is included; see Fig. 1 in Ref. \cite{suv20}) for the Brans-Dicke choices discussed in Sec. III. B; i.e., $F = \phi$, $\mathcal{V}=0$, and $\chi = \chi_{0}/\phi$ for some positive $\chi_{0}$. Solving for the scalar field in the constraint equation \eqref{eq:constraint} with the property that $X_{0} = 0$, i.e.,
\begin{equation} \label{eq:partcons}
0 = \pb(r) \Rb - \chi_{0} \frac {\pb'(r)^2} { \pb(r) B(r)},
\end{equation} 
leads to an exact solution in the mixed scalar-tensor theory \eqref{eq:genscalarten} with $f(X) = X^{1+\sigma}$ for any $\sigma > 0$, as discussed in Sec. II. A. In general, one should impose the boundary condition 
\begin{equation}
\underset{r \rightarrow \infty}{\lim} \pb(r) = 1,
\end{equation}
so that the scalar field approaches the Newtonian value, viz. $F(\phi) \rightarrow 1$. Equation \eqref{eq:partcons} informs us that not all arbitrary combinations of the deformation parameters ($\epsilon, a_{1}, b_{1}$) are permitted within the aforementioned theory. Some combinations can lead to ghost-like instabilities where $\phi \leq 0$ (i.e., non-positive rest energy) or $\phi'(r)^2 < 0$ (i.e., negative kinetic energy), and care must be taken to ensure that such undesirable features do not emerge \cite{noj19,felice10}. For all parameter combinations considered here, the scalar fields are positive, smooth, and have non-negative kinetic energy.

\begin{figure}
	\centering
	\includegraphics[width=1.0\linewidth]{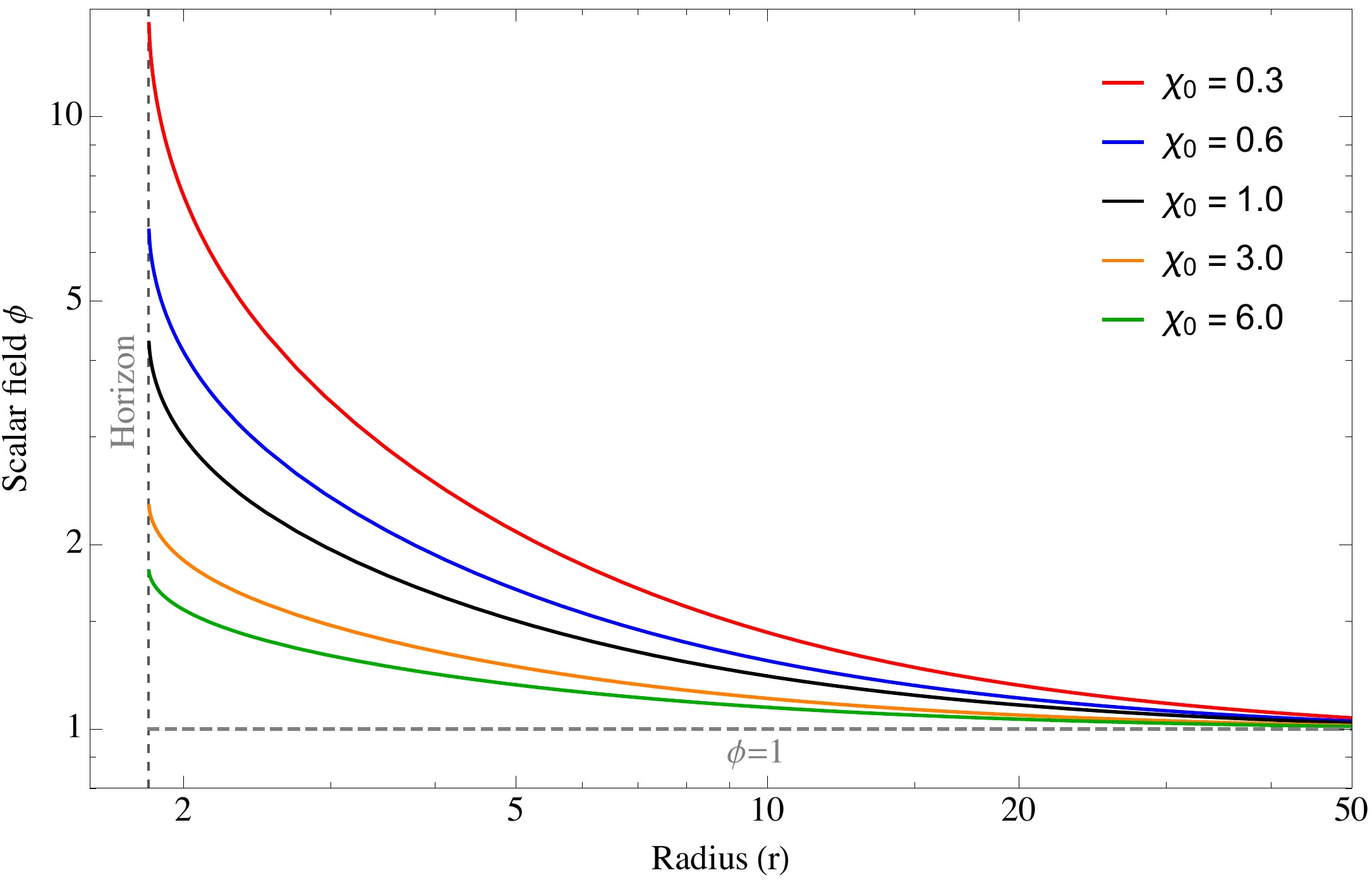}
	\caption{Radial scalar field profiles as solutions to equation \eqref{eq:partcons} for a variety of $\chi_{0}$ values (see plot legends), where we take $M=1$, $\epsilon = 0.1$, $a_{1} = 0.15$, and $b_{1} = - 0.4$. The horizon, occurring at $r = \rH = 20/11$, is shown by a vertical dashed line. }
	\label{scalarfields}
\end{figure}

Figure \ref{scalarfields} shows a sample of radial scalar field profiles as solutions to equation \eqref{eq:partcons} for a variety of $\chi_{0}$ values and a representative set of RZ parameters (henceforth, we set $M=1$ throughout the rest of the work): $\epsilon = 0.1$, $a_{1} = 0.15$, and $b_{1} = - 0.4$. We see that the amplitude of the scalar field is directly proportional to the reciprocal of $\chi_{0}$ and furthermore monotonically decreases as a function of radius, with maxima and minima attained at the horizon and infinity, respectively. For instance, the $\chi_{0} = 0.3$ (red curve) scalar field solution is $\gtrsim 10$ times larger than the $\chi_{0} = 6$ (green curve) case near the horizon, though at $r \sim 50$ the two agree to within a few percent. The scalar hair is short-ranged in all cases therefore, and would be virtually invisible at large distances from the horizon even for relatively small values of $\chi_{0}$. Given a solution to equation \eqref{eq:partcons}, one can now proceed to evaluate the axial potentials $V_{\ell}$, as necessary to compute QNMs.

\subsection{Numerical methods}

In general, the Schr{\"o}dinger-like equation \eqref{eq:rweqn} is subject to some set of boundary conditions. These are chosen such that we have pure outgoing radiation at infinity and purely ingoing radiation at the horizon, i.e., that as $r^{\star} \rightarrow \pm \infty$, we have \citep{chand75}
\begin{equation} \label{eq:boundarycond}
Z(\omega,r) \sim e^{\pm i \omega r^{\star}}.
\end{equation}
Equation \eqref{eq:rweqn}, subject to the boundary conditions \eqref{eq:boundarycond}, constitutes an eigenvalue problem for the QNMs. However, since we are mostly interested in the frequencies of the QNMs and not so much the functional form of $Z$ itself, it is not necessary to solve equation \eqref{eq:rweqn} formally (see Refs.~\cite{Kokkotas:1999bd,Nollert:1999ji,Berti:2009kk} for some classical reviews on the topic).

Although there are a number of semi-analytic and numerical methods available, we operate with the WKB method  \cite{Schutz:1985km,Iyer:1986np,Konoplya:2003ii} because it is especially suited for problems in which other standard approaches may require significant adjustments or extensions \cite{Konoplya:2006rv}. Other techniques involving continued fractions (e.g., Leaver's method \cite{Leaver:1985ax}), phase-integrals (e.g., \cite{and92}), or time-dependent integration (e.g., \cite{Chandrasekhar:1975zza}) typically require an analytic understanding of the asymptotic properties of the potential, which may be absent when one only has numerical potentials defined up to some finite radius at their disposal, as in our case. In the order-$j$ WKB method, the QNM frequency is determined through
\begin{align}\label{eq:wkb-expansion}
\frac{i Q_0}{\sqrt{2 Q^{\prime \prime}_0}} - \sum_{j=2}^{} \Lambda_j(n) = n + \frac{1}{2},
\end{align}
for overtone number\footnote{Throughout this work we only work with the least-damped (and therefore most relevant for astrophysical observation) modes for brevity, which have $n=0$ by definition. However, overtones are generally expected to be active for a newborn BH and may be important when reconstructing BH parameters from astrophysical data; see Ref. \cite{gies19} for a discussion.} $n$, where $Q(r^{\star}) \equiv \omega_n^2 - V_{\ell}(r^{\star})$ is evaluated at the maximum of the potential and primes denote derivatives with respect to the tortoise coordinate $r^{\star}$. The $\Lambda_j$ in equation \eqref{eq:wkb-expansion} depend on $2j^{\text{th}}$-order derivatives of $V_{\ell}$ and have lengthy forms that we will not write out here; see, e.g., Ref. \cite{Konoplya:2003ii}. 

While the WKB method has been extended up to 13th order using Pad\'e approximants [thereby requiring $26^{\text{th}}$-order derivatives of $V_{\ell}(r)$] \cite{Matyjasek:2017psv}, numerical stability and implementation considerations limit us to the $4^{\text{th}}$-order scheme. The relative errors of our results have been checked using both $3^{\text{rd}}$ and $4^{\text{th}}$ order approximations and are, at worst, at the percent level for a few cases (imaginary parts). For most cases however the errors are significantly smaller. An additional numerical check has been performed using the P\"oschl-Teller approximation \cite{Ferrari:1984ozr,Ferrari:1984zz} to verify the robustness of the WKB method. Note that for a few parameter combinations the $\ell =2$ potentials develop a small negative gap left of the maximum (see Figure \ref{pot1} below), which is likely responsible for the percent-level errors mentioned previously.  In any case, the various numerical checks performed in this work make us confident that the numerical routines for finding the potential [checking that the right-hand side of \eqref{eq:partcons} vanishes to machine precision] and computing the QNMs (checking agreement between $3^{\text{rd}}$ and $4^{\text{th}}$ orders and the P\"oschl-Teller approximation) give reliable results.

We graph the $\ell=2$ potential profiles $V_{0,2}(r)$ (solid curves) for two illustrative cases with $a_{1} = b_{1} = 0$ and $M = \chi_{0} =1$ but $\epsilon = -0.1$ (red) and $\epsilon = -0.3$ (blue) in Fig. \ref{pot1}. For contrast, the corresponding potentials computed using the GR-like scheme $[\pb'(r) = 0]$ are represented by dashed curves. The respective domains of the potential functions vary with $\epsilon$, because the horizon location $\rH$ (shown by vertical dashed lines) is sensitive to this quantity. In the case of low $|\epsilon|$ (i.e., more Schwarzschild-like), we see that the GR-RZ and exact potential functions overlap almost exactly: the greatest difference between the two curves is $\sim 1.5\%$ near the respective peaks (shown by solid diamonds). However, since the properties of the QNMs are directly tied to the location and value of the peak [cf. equation \eqref{eq:wkb-expansion}], even a small difference can lead to a non-trivial disparity in the real and imaginary components (see Sec. IV. B). For the $\epsilon = -0.3$ case, the difference between the exact and approximate potentials is more noticeable, and the relative difference reaches $\sim 7\%$ near the respective peaks. Note that a negative gap develops near the horizon in the exact potential for the large $|\epsilon|$ case, though is not deep enough to produce bound states for the chosen kinetic value $\chi_{0} = 1$. For $\chi_{0} \ll 1$ or $\epsilon \ll -0.3$, bound states may exist in principle. More generally, the disagreement between the GR and exact schemes scales inversely with $\chi_{0}$ and directly with $|\epsilon|$ (see Figures \ref{qnm_chi} and \ref{qnm_eps} below, respectively).

\begin{figure}
	\centering
	\includegraphics[width=1.0\linewidth]{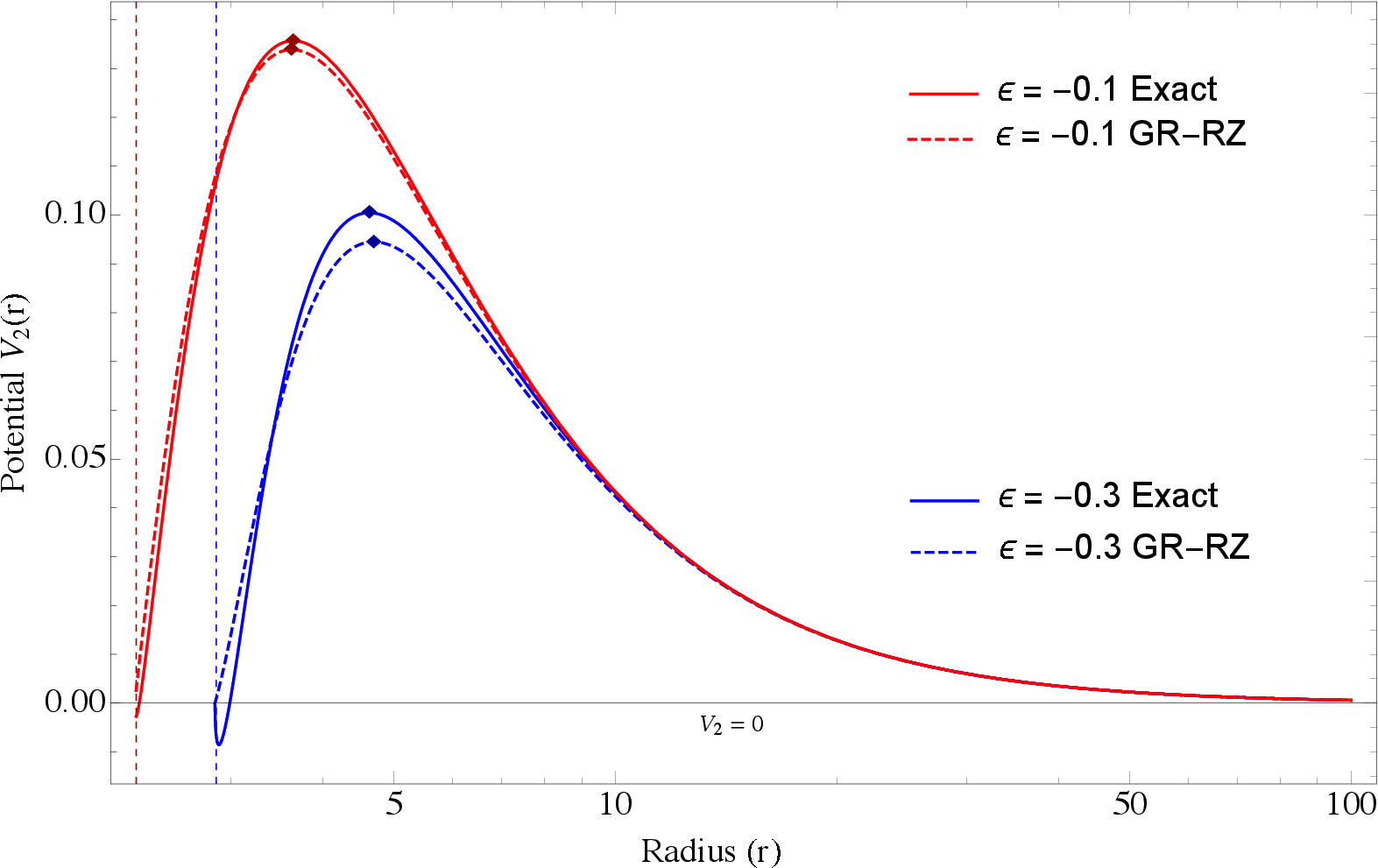}
	\caption{Comparison between the $\ell=2$ exact [$V_{0,2}$ from equation \eqref{eq:vsing}; solid curves] and GR-like (dashed curve) potential profiles for $a_{1} = b_{1} = 0$ and $M = \chi_{0} =1$ but $\epsilon = -0.1$ (red curves) and $\epsilon = -0.3$ (blue curves). The location of the respective horizons are shown by vertical dashed lines. The location of the maxima for the peaks, relevant for the computation of QNMs, are shown by solid diamonds. A negative gap forms in the exact potential for $\epsilon \lesssim -0.3$. }
	\label{pot1}
\end{figure}

\subsection{Results}

In this section we present computations of QNMs using the numerical methods detailed in the previous section for a variety of representative cases.

\begin{figure}
	\centering
	\includegraphics[width=1.0\linewidth]{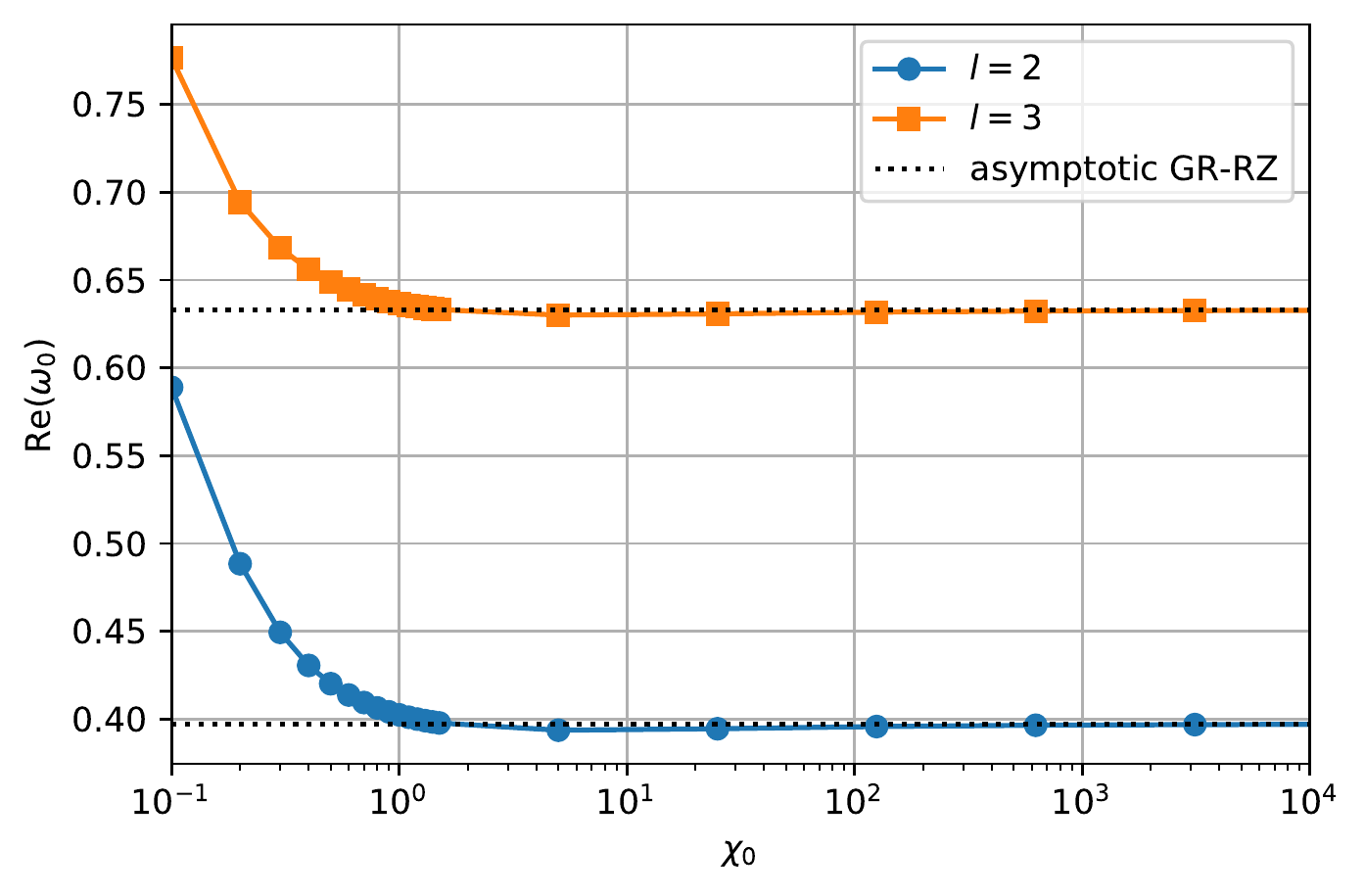}
	\includegraphics[width=1.0\linewidth]{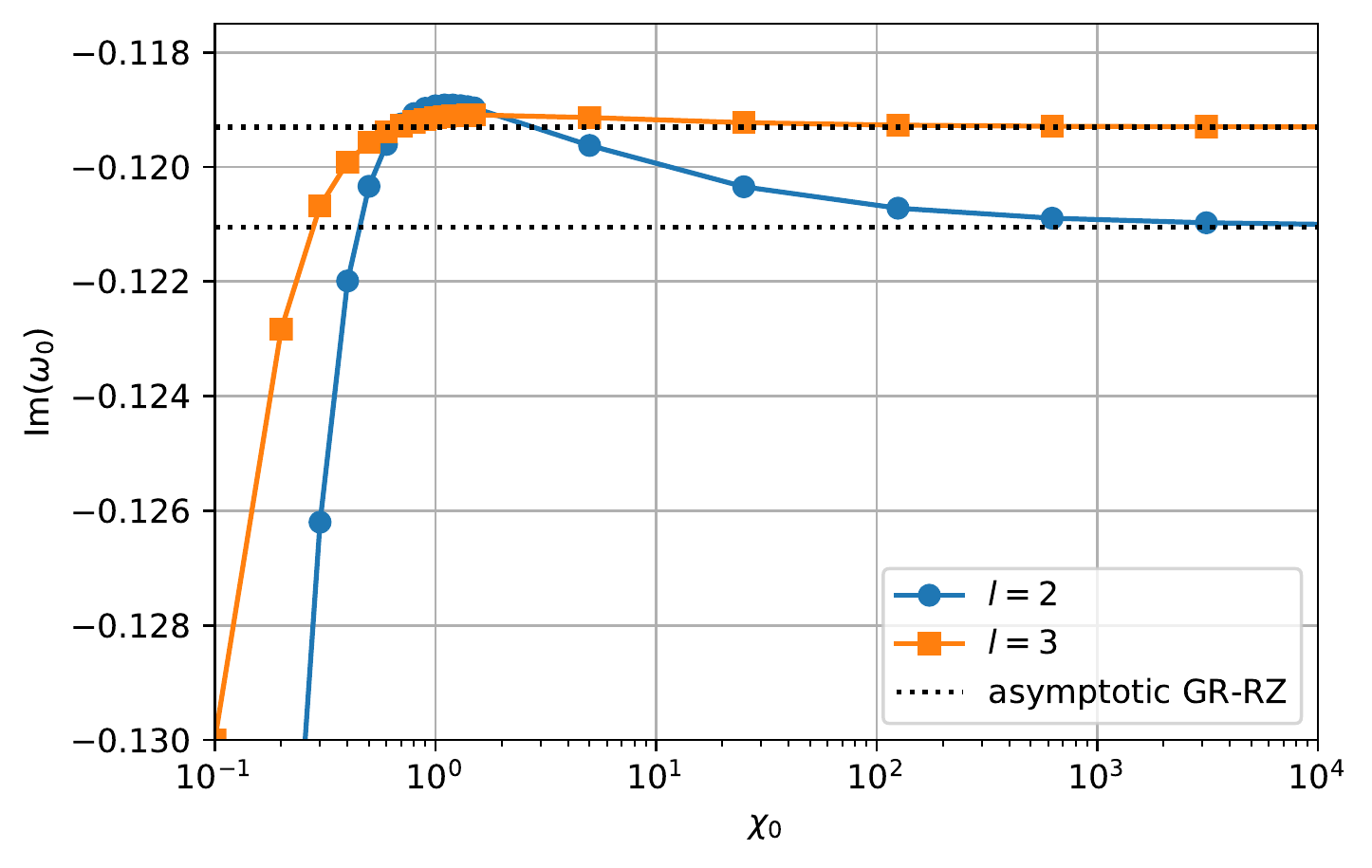}
	\caption{Fundamental $(n=0)$ QNM frequencies $\omega_0$ as functions of $\chi_{0}$ for $\ell =2$ (blue circles) and $\ell =3$ (orange squares), where we fix $\epsilon=0.1, a_1=0.15$, and $b_1=-0.4$. The upper panel displays the real part of $\omega_0$, and the lower panel the imaginary part. Overplotted (black dots) are the frequencies obtained when using the GR-like approximation scheme detailed in the text.}
	\label{qnm_chi}
\end{figure}

Fig. \ref{qnm_chi} shows the $\ell =2$ (blue circles) and $\ell = 3$ (orange squares) QNM frequencies as functions of $\chi_{0}$ for the RZ parameters relevant to the scalar field solutions shown in Fig. \ref{scalarfields} (i.e., $\epsilon = 0.1, a_{1} = 0.15$, and $b_{1} = -0.4$). For small values of $\chi_{0}$ in the range $0.1 \leq \chi_{0} \lesssim 1.5$, we see a substantial and monotonic variation in both the real ($\approx 33\%)$ and imaginary ($\approx 15\%$) components of the $\ell =2$ QNMs. Variations on a similar scale are likewise observed in the $\ell = 3$ case. Turning points occur in the graphs once a critical value of $\chi_{0} \approx 1.5$ is reached, and the real (imaginary) frequencies begin to increase (decrease) rather than decrease (increase). In the large $\chi_{0}$ limit, both sets of frequencies asymptote towards particular GR-like values (see below). In any case, we find that a continuous (with respect to $\chi_{0}$) spectrum of frequencies within some bounded range can be achieved for a fixed set of RZ parameters. This result has the interesting implication that, given only a single measurement of the fundamental frequency of a newborn object (as in the case of GW150914 \cite{gw150914}), many RZ parameter sets can be accommodated within \emph{some}\footnote{Note, however, it may be the case that that particular branch {is Ostrogradsky-unstable \cite{wood07} or} does not respect independent constraints coming from cosmology \cite{bohm08,sharm20}, neutron-star astrophysics \cite{cap16}, or Solar-system dynamics \cite{will18}.} branch of the general theory \eqref{eq:genscalarten}. This stresses the necessity of having multiple QNM measurements in placing constraints on BH behaviour and strong gravity simultaneously (see also Refs. \cite{Volkel:2019muj,mcm1,mcm2}).

Overplotted in Fig. \ref{qnm_chi} are the GR-like QNMs (black dots), which of course do not vary as a function of $\chi_{0}$. For small of values of $\chi_{0}$, the two schemes predict distinct behaviour, as noted previously. In the limit $\chi_{0} \rightarrow \infty$ however, the scalar dynamics are heavily suppressed (cf. Fig. \ref{scalarfields}) and the curves match as the theory approaches the Buchdahl one \cite{buch70}, much in the same way that the Brans-Dicke theory (i.e., linear $f$) approaches GR in this limit. In practice, values $\chi_{0} \gtrsim 10^{4}$ lead to numerical indistinguishability between the real and imaginary components for the GR-like and exact schemes for this particular set of RZ parameters. Note though that this does not mean that the frequencies approach the Schwarzschild values: the $\ell=2$ fundamental mode in the Schwarzschild case (not shown) has real value $\text{Re}(\omega_{\text{GR}}) = 0.374$ \cite{Chandrasekhar:1975zza}, for instance, and is still $\sim 8\%$ different from the RZ value in this particular case even when $\chi_{0} \rightarrow \infty$. Such a disparity does not, however, exceed the limits imposed by observations of GW150914, where constraints on the fundamental frequency were placed at roughly the $\sim 10\%$ level relative to the GR (Kerr) values at $90\%$ confidence \cite{gw150914}.

\begin{figure}
	\centering
	\includegraphics[width=1.0\linewidth]{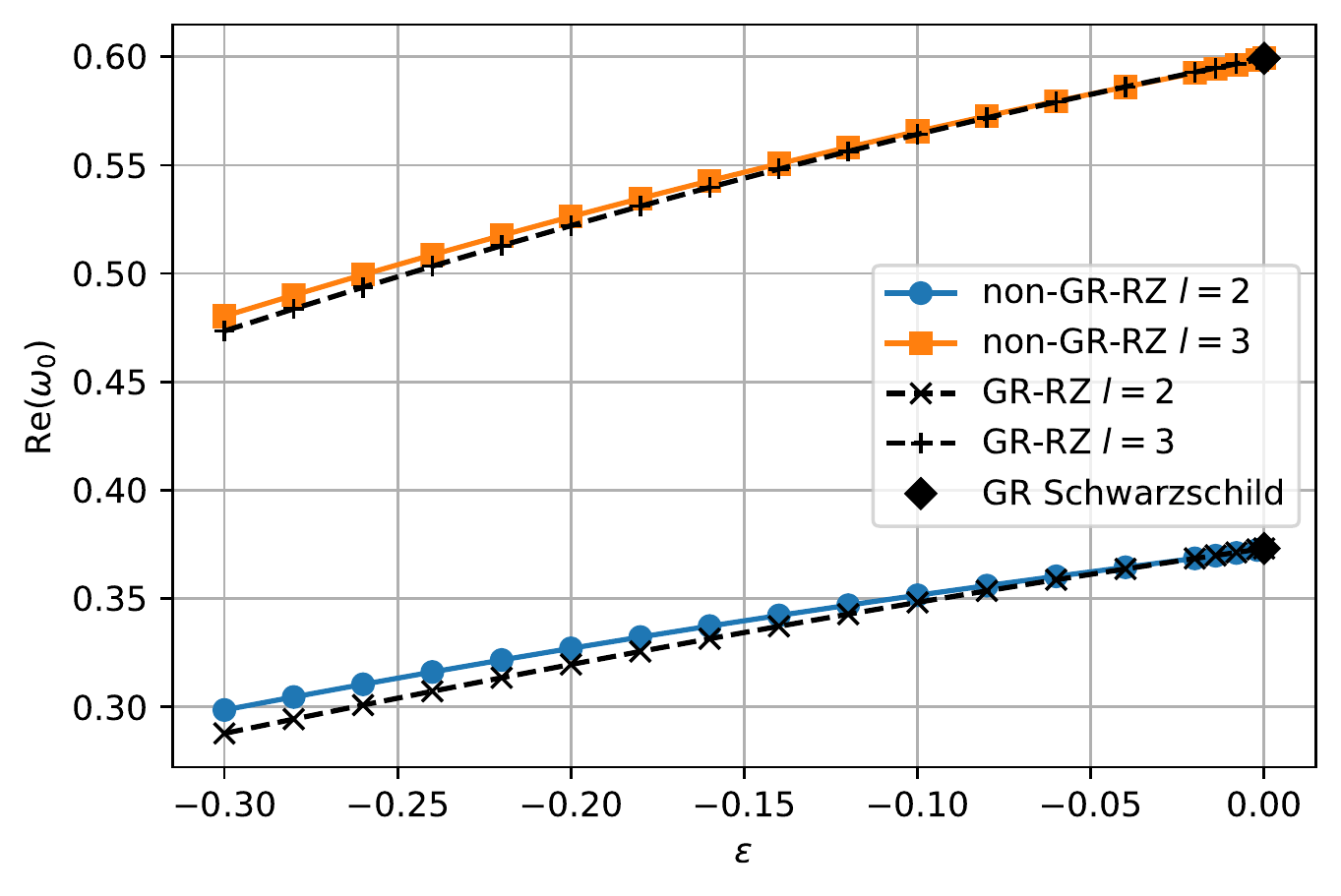}
	\includegraphics[width=1.0\linewidth]{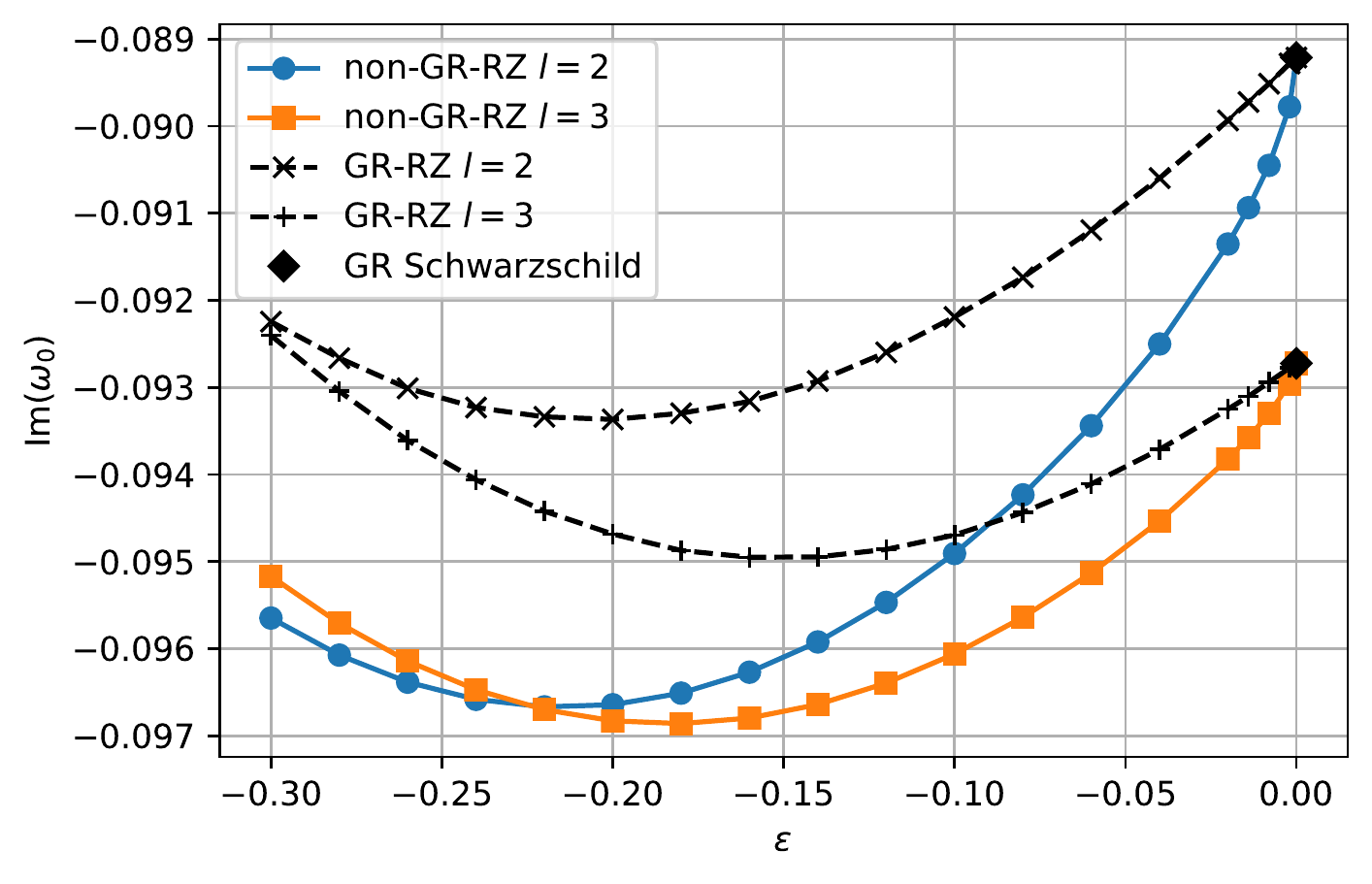}
	\caption{Similar to Fig. \ref{qnm_chi} though displaying $\omega_{0}$ as a function of $\epsilon$ for fixed values $a_{1} = b_{1} = 0$ and $\chi_{0} = 1$.}
	\label{qnm_eps}
\end{figure}

As another example, we consider the case of $a_{1} = b_{1} = 0$ with fixed $\chi_{0} = 1$ but varying $\epsilon$. This case is illustrated in Fig. \ref{qnm_eps} in a similar style to Fig. \ref{qnm_chi}. In this instance, the GR-like scheme is approached as $\epsilon \rightarrow 0$, where in fact the Schwarzschild QNMs are recovered exactly \cite{Chandrasekhar:1975zza} in both schemes, as expected. Overall, we see that the GR-like approximation is a robust one: disagreements, relative to the exact case, in the real and imaginary parts of the QNM frequencies are at most $\sim 3\%$ for $|\epsilon| \lesssim 0.3$ and $\ell =2$. For larger values of $\chi_{0}$ and $\ell$, the disagreements fall even further. This adds strength to the claims made in Ref. \cite{Volkel:2020daa}, who made use of the GR-like approximation within a Bayesian scheme to show how QNM data can be used to reconstruct a spacetime metric to a high degree of accuracy. In either scheme, however, we see that between the $\epsilon = 0$ (Schwarzschild value) and $\epsilon = -0.3$ cases, the real parts of the QNM frequencies change by $\approx 20\%$, which exceeds the limits imposed by GW150914 \cite{gw150914}. Demanding that the frequencies match to the Schwarzschild values within $\lesssim 10\%$ leads to the constraint $|\epsilon| \lesssim 0.16$ for $\chi_{0} =1$, though we note that such a direct comparison with GW data is imprecise at this stage because we do not model rotation.

Although the parameter space of RZ-QNMs is very large (different quantum numbers, RZ parameters, and theory variables) and an exhaustive study is impractical, it is instructive to consider a few additional cases. The $\ell \leq 5$ QNM frequencies for a few cases with scattered values of $\epsilon$, $a_{1}$, and $b_{1}$ are shown in Table \ref{tab:qnm1} (\ref{tab:qnm5}) for  $\chi_{0} = 1$ ($\chi_{0} = 0.5$). For example, the second and third columns list the frequencies for cases where all RZ parameters vanish except for $b_{1}$. The real parts of the frequencies are largely similar to the Schwarzschild values in these cases for $b_{1} = -0.38$, where we find $\text{Re}(\omega) = 0.38$ for $\chi_{0} = 1$ while $\text{Re}(\omega) = 0.368$ for $\chi_{0} = 0.5$, which are marginally larger and smaller than the Schwarzschild value $\text{Re}(\omega_{\text{GR}}) = 0.374$, respectively. The imaginary components are substantially greater in either case ($\gtrsim 40\%$ difference) though, implying that the modes would be damped out faster [since the damping time $\propto -\text{Im}(\omega)^{-1}$] than the corresponding GR case. For more negative values of $b_{1}$, however, even the real values diverge significantly from the Schwarzschild values, and are thus likely ruled-out from GW observations \cite{ligo19}. The damping times for the cases shown in the final two columns, which have $\epsilon = 0.1$ and $b_{1} = -0.19$ but $a_{1} = 0.15$ and $a_{1} = 0.3$, respectively, are very similar to each other, and only weakly depend on $\ell$. This implies that even high-$\ell$ modes may be important in the early characterisation of BH ringdown for some non-Schwarzschild metrics, since they may not necessarily be damped out faster than their low-$\ell$ counterparts.

\begin{table*}
\caption{Selected $\ell \leq 5$ QNM frequencies for a few RZ parameter combinations with $\chi = 1/\phi$ (i.e., $\chi_{0} = 1$). The columns headed by vectors $(q_{1},q_{2},q_{3})$ refer to the values of the RZ parameters: $\epsilon = q_{1}$, $a_{1} = q_{2}$, and $b_{1} = q_{3}$, respectively. The complex frequencies are given in geometrical units, though can be converted into physical units through a multiplication by $\left(2 \pi \times 5142 \times M_{\odot}/M \right) \text{ Hz}$ (see, e.g., Ref. \cite{Kokkotas:1999bd}). For reference, the Schwarzschild values are given by $\omega_{\text{GR}} = 0.374 - 0.089 i$ for $\ell =2$ and $\omega_{\text{GR}} = 0.599 - 0.093 i$ for $\ell =3$.}
\centering
  \begin{tabular}{|l|c|c|c|c|c|}
  \hline
  \hline
Quantum number &  (0,0,-0.38) & (0,0,-0.57)  & (0.1,0.15,-0.19) & (0.1,0.3,-0.19)  \\
\hline
$\ell = 2 $ & $ 0.380 -0.127 {i} $ & $ 0.484 -0.162 {i} $ & $ 0.395 -0.092 {i} $ & $ 0.438 -0.119 {i}  $   \\
$\ell = 3 $ & $ 0.597 -0.123 {i} $ & $ 0.677 -0.144 {i} $ & $ 0.634 -0.104 {i} $ & $ 0.677 -0.116 {i}  $   \\
$\ell = 4 $ & $ 0.805 -0.120 {i} $ & $ 0.867 -0.136 {i} $ & $ 0.856 -0.104 {i} $ & $ 0.900 -0.114 {i}  $   \\
$\ell = 5 $ & $ 1.008 -0.119 {i} $ & $ 1.059 -0.133 {i} $ & $ 1.072 -0.105 {i} $ & $ 1.118 -0.113 {i}  $   \\
\hline
\hline
\end{tabular}
\label{tab:qnm1}
\end{table*}

\begin{table*}
\caption{Similar to Tab. \ref{tab:qnm1} but for $\chi(\phi) = 1/(2 \phi)$, i.e., $\chi_{0} = 0.5$.}
\centering
  \begin{tabular}{|l|c|c|c|c|c|}
  \hline
  \hline
Quantum number &  (0,0,-0.38) & (0,0,-0.57)  & (0.1,0.5,-0.19) & (0.1,0.3,-0.19)   \\
\hline
$\ell = 2 $ & $ 0.368 -0.123 {i} $ & $ 0.420 -0.142 {i} $ & $ 0.394 -0.097 {i} $ & $ 0.420 -0.112 {i}  $   \\
$\ell = 3 $ & $ 0.588 -0.120 {i} $ & $ 0.628 -0.136 {i} $ & $ 0.633 -0.104 {i} $ & $ 0.664 -0.114 {i}  $   \\
$\ell = 4 $ & $ 0.798 -0.118 {i} $ & $ 0.829 -0.133 {i} $ & $ 0.856 -0.104 {i} $ & $ 0.890 -0.114 {i}  $   \\
$\ell = 5 $ & $ 1.002 -0.117 {i} $ & $ 1.028 -0.132 {i} $ & $ 1.072 -0.105 {i} $ & $ 1.109 -0.113 {i}  $  \\
\hline
\hline
\end{tabular}
\label{tab:qnm5}
\end{table*}

\section{Discussion}

Recently, a solution to the gravitational inverse problem was presented in Ref. \cite{suv20}: given some metric $g$, a covariant, scalar-tensor theory of gravity [with action \eqref{eq:genscalarten}] can be designed such that that particular $g$ is an exact to the \emph{vacuum} field equations \eqref{eq:field1} and \eqref{eq:field2} (see Sec. II. A). A practical application of this result is that bottom-up BH metrics can be assigned to an exact theory of gravity, which allows for a self-consistent study of their perturbations. In this work, we derive the EOM describing axial perturbations of static, spherically symmetric spacetimes in this theory (Sec. III).  As a demonstration of the mathematical machinery, we compute the QNMs for the RZ metric for a variety of non-GR parameters using WKB methods (Sec. IV). The approach presented here is not unique to the RZ metric, and the method described here can be readily adapted to practically any class of static, parameterised BH metrics.

While realistic BHs rotate in reality, understanding the QNM spectrum of static objects is still useful in characterising hypothetical signatures of modified gravity in the strong-field regime. In particular, V{\"o}lkel and Barausse \cite{Volkel:2020daa} have shown how ringdown data can be used to reconstruct the local spacetime metric given a theory of gravity (see also Refs. \cite{mcm1,mcm2,Volkel:2019muj}). In that work, however, a solution to the inverse problem was not available, and so various approximations for the EOM describing QNMs had to be used. This work may therefore help to maximise the information gleaned from future GW measurements when combined with statistical analyses along the lines presented in Ref. \cite{Volkel:2020daa}. For large values of the Brans-Dicke parameter $\chi_{0}$, however, this extra step may not be necessary since we found that the exact results are well-approximated by the GR-like scheme. 

At present, constraints on departures from the GR fundamental frequency are at roughly the $\sim 10\%$ level at 90$\%$ confidence \cite{gw150914}. For the RZ metric specifically, we find that this corresponds to a constraint $|\epsilon| \lesssim 0.16$ for $\chi_{0} =1$ when $a_{1} = b_{1} = 0$, as can be seen from Fig. \ref{qnm_eps}. There is of course additional uncertainty since the QNM frequencies scale with $\chi_{0}$ and the other RZ parameters too; see Fig. \ref{qnm_chi} and Tabs. \ref{tab:qnm1} and \ref{tab:qnm5}. Overall, we validate the results of Ref. \cite{Volkel:2020daa} where GR-like perturbation equations were used, and find that a non-trivial RZ parameter space is consistent with current ringdown bounds.

There are several directions in which extensions of this work would be worthwhile. One of these is to include rotation, as mentioned previously: the difficulty in this is largely computational in nature, since the perturbation equations are in general coupled and solving the associated eigenvalue problem requires more involved techniques, such as generalizations of Leaver's method \cite{Leaver:1985ax} or direct time domain computations. In particular, the inverse problem as presented here is still relatively straightforward to handle for stationary spacetimes (see the example given in Ref. \cite{suv20}), and so this aspect of the work is not difficult to extend. In a similar way, we have only looked at axial QNMs here, though polar perturbations are expected to carry $\sim 50\%$ of the GW energy away from a newborn BH due to Regge-Wheeler-Zerilli isospectrality (though cf. Refs. \cite{bhatt17,dat20}), and are therefore astrophysically important. Further investigation of the existence of bound states for non-RZ BHs using the S-deformation technique associated with equation \eqref{eq:sdef} is also interesting, since it is known that even Kerr black holes can be unstable in some theories \cite{card09,gao19}. Using the methods presented here, one could attempt to map out the space of stable BH solutions in a theory-dependent manner (e.g., by considering stability as a function of $\chi_{0}$). Finally, the scalar-tensor class of theories \eqref{eq:genscalarten} is not the only type of theory that can be designed to solve the inverse problem. Mixed vector-$f(R)$ theories (e.g., generalised Proca theories \cite{hei14}) also provide examples of solutions to the inverse problem; see Ref. \cite{suv20} for a discussion. It would be worthwhile to study the QNM spectrum of RZ or other black holes in these theories in future.

\section*{Acknowledgments}
AGS is supported by the Alexander von Humboldt Foundation. SV acknowledges financial support provided under the European Union's H2020 ERC Consolidator Grant ``GRavity from Astrophysical to Microscopic Scales'' grant agreement no. GRAMS-815673. SV thanks Enrico Barausse, Marco Crisostomi, and Roman Konoplya for useful discussions.

\bibliography{literature}

\end{document}